\documentclass[runningheads]{llncs}
\usepackage[T1]{fontenc}
\usepackage{graphicx}
\usepackage{float}
\usepackage{soul}
\usepackage[colorlinks, citecolor=blue]{hyperref}
\usepackage{amsmath}
\usepackage{booktabs}
\usepackage{subcaption}
\usepackage{parskip}
\usepackage{cite}
\usepackage{amsmath,amssymb,amsfonts}
\usepackage{algorithmic}
\usepackage{textcomp}
\usepackage{xcolor}
\def\BibTeX{{\rm B\kern-.05em{\sc i\kern-.025em b}\kern-.08em
    T\kern-.1667em\lower.7ex\hbox{E}\kern-.125emX}}
\usepackage{graphicx}

\begin{document}

\title{Wine feature importance and quality prediction: A comparative study of machine learning algorithms with unbalanced data}

\titlerunning{Wine features importance and quality prediction}

\author{Siphendulwe Zaza\inst{1}\email{zazasiphendulwe@gmail.com} \\
Marcellin Atemkeng\inst{1*}\email{m.atemkeng@ru.ac.za} \\
Sisipho Hamlomo\inst{1,2}\email{s.hamlomo@ru.ac.za}}

\authorrunning{Siphendulwe Zaza, Marcellin Atemkeng, and Sisipho Hamlomo}

\institute{Department of Mathematics, Rhodes University, Grahamstown,  South Africa \and
Department of Statistics, Rhodes University, Grahamstown, South Africa}

\maketitle

\begin{abstract}
Classifying wine as "good" is a challenging task due to the absence of a clear criterion. Nevertheless, an accurate prediction of wine quality can be valuable in the certification phase. Previously, wine quality was evaluated solely by human experts, but with the advent of machine learning this evaluation process can now be automated, thereby reducing the time and effort required from experts. The feature selection process can be utilized to examine the impact of analytical tests on wine quality. If it is established that specific input variables have a significant effect on predicting wine quality, this information can be employed to enhance the production process. We studied the feature importance, which allowed us to explore various factors that affect the quality of the wine. The feature importance analysis suggests that alcohol significantly impacts wine quality. Furthermore, several machine learning models are compared, including Random Forest (RF), Support Vector Machine (SVM), Gradient Boosting (GB), K-Nearest Neighbors (KNN), and Decision Tree (DT). The analysis revealed that SVM excelled above all other models with a 96\% accuracy rate.
\end{abstract}
\section{Introduction}
The quality of wine is very important for both consumers and the wine industry therefore, it is imperative to determine wine quality before manufacturing or consumption. However, relying on human expert wine tasting for measuring wine quality can be a time-consuming and subjective process, posing significant challenges for experts in providing accurate predictions. According to \cite{adam2009health}, wine testing by human experts can also put them at health risk as they are exposed to a range of chemicals and other substances that may be harmful to their health. For example, the inhalation of volatile organic compounds (VOCs) such as ethanol, acetaldehyde, and ethyl acetate, during the process of wine tasting has been linked to a range of health issues, including headaches, coughs, and respiratory problems \cite{saremi2008cardiovascular, Meyer2019}. With the aid of machine learning algorithms, it is now possible to analyze the physiochemical properties of wine, which can be used to predict its quality. The aim of this paper is to use the chemical and physical properties of wine to predict its quality and to determine which features are more important for predicting good wine. We use the following algorithms: Decision Tree (DT), Random Forest (RF), Support Vector Machine (SVM), K-Nearest Neighbors (KNN), and Gradient Boosting (GB). These models are used due to the nature of the wine data we used to run the experiment. The data is of small samples, and it is also imbalanced. Shallow machine learning models have shown the potential to outperform deep learning models on small datasets. For example, \cite{dahal2021prediction} and \cite{Dua:2019} used some of the above-mentioned shallow machine learning models on small datasets, and these algorithms have shown exceptional performance in addressing the challenges of small sample sizes and imbalanced data.

The contribution of this work is as follows: we have trained five models and compared their performance on an unbalanced dataset, then we move further to use some sampling methods to balance the dataset and then retrain the models. Sampling methods improved the accuracy of the models with SVM resulting from 78\% without sampling to 96\% with sampling, thereby outperforming other models.

\section{Related Work}
\cite{gupta2018selection} has employed a range of machine learning techniques such as linear regression to find important features for prediction and also used SVM and neural networks to predict values. The conclusion is reached that not all features are important for predicting wine quality hence one can select features that are most likely to be useful for predicting the quality of the wine. They used both the white wine and red wine datasets for their analysis, which is slightly different from our work. In our study, we focused only on the red wine dataset for our analysis and we compared our study with the work of \cite{gupta2018selection} who used two datasets which are the white wine and red wine datasets. Our findings with the red wine dataset aligned with the results in \cite{gupta2018selection} for predicting wine quality.

\cite{pawar2019wine} employed four machine learning techniques namely RF, stochastic gradient descent, SVM, and logistic regression to forecast the quality of the wine. Out of the four techniques, RF outperformed other methods with an accuracy of 88\%. In the latter work, the red wine dataset is used \cite{Dua:2019}, which was then divided into two classes namely good wine and bad wine. Our research is similar to this, but we attempted to extend the problem by introducing three classes. We found that SVM was the best-performing model for predicting the quality of wine, with an accuracy of 96\% compared to the 88\% accuracy achieved by RF in \cite{pawar2019wine}.
In \cite{dahal2021prediction} the naive Bayes, DT, SVM, and RF are used to predict wine quality. The analysis shows that when the residual sugar is minimal the quality of the wine increases and does not change significantly, suggesting that this feature is not as important as others such as alcohol and citric acid. We also observed in the research that our machine learning models were producing acceptable results when residual sugar was excluded. This suggests that residual sugar is not an important feature when predicting wine quality.
\vspace{-0.4cm}
\section{Data description and preprocessing}
\vspace{-0.2cm}
\subsection{Data description}
The red wine dataset utilized in this study is sourced from the UCI machine learning repository \cite{cortez2009modeling}. This dataset comprises 1599 instances of red wine, and its quality is assessed through 11 distinct input variables including Fixed acidity, Volatile acidity, Citric acid, Residual sugar, Chlorides, Free sulfur dioxide, Total sulfur dioxide, Density, PH, Sulphates, and Alcohol. The output variable quality is based on these input parameters and is rated on a scale of 0 to 10, with 0 representing poor wine and 10 signifying excellent wine. Table  \ref{tafile1} presents the statistical summary of the red wine dataset employed in this paper.

\begin{table}[H]
   \centering
   \begin{tabular}{|c|c|c|c|c|c|}
   \hline
   		\textbf{Variable Name} & \textbf{Mean} & \textbf{Sd} & \textbf{Min} & \textbf{Max} & \textbf{Median}\\
   		\hline\hline
   		Fixed acidity & 8.31 & 1.73 & 4.60 & 15.90 & 7.90 \\
   		Volatile acidity & 0.52& 0.18& 0.12& 1.58& 0.52\\
   		Citric acid & 0.27 & 0.19& 0.00& 1.00& 0.26\\
   		Residual sugar& 2.52& 1.35& 0.90& 15.50&2.20\\
   		Chlorides& 0.08& 0.04& 0.01& 0.61&0.07\\
   		Free sulfur dioxide& 15.89& 10.44& 1.00& 72.00& 14.00\\
   		Total sulfur dioxide& 46.82& 33.40& 6.00& 289.00& 38.00\\
   		Density& 0.99& 0.001& 0.99& 1.00& 0.99\\
   		PH& 3.30& 0.15& 2.74& 4.01& 3.31\\
   		Sulphates& 0.65& 0.17& 0.33& 2.00& 0.62\\
   		Alcohol& 10.43& 1.08& 8.40& 14.90& 10.20\\
   		Quality& 5.62& 0.82& 3.00& 8.00&8.00\\
   		\hline
   \end{tabular}
   \caption{statistics for red wine dataset}
   \label{tafile1}
\end{table}

\subsection{Data Pre-processing}
\begin{figure}
\begin{center}
\includegraphics[width=8cm, height=6cm]{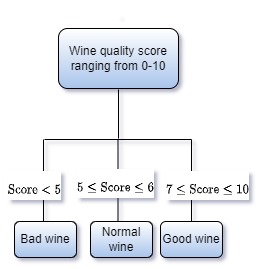}
\end{center}
\caption{Label encoding}
\label{flow}
\end{figure}

We use label encoding, a process that converts the labels into a machine-readable form. We use this method to categorize the data into good, normal, or bad categories. We label bad wine as wine with a quality score that is less than 5, normal wine as wine with a quality score that is between 5 and 6, and good wine as wine with a quality score between 7 and 10, as shown in the flowchart in Figure \ref{flow}. Also as part of data pre-processing, we excluded duplicate entries and data points with missing values in the dataset to maintain the integrity of the analysis. 

\subsection{Data analysis}
The covariance matrix provides values within the range of $(-1,1)$ which gives us information about the relationship between variables. A value of 1 indicates a strong positive linear correlation between variables whereas -1 suggests a strong negative linear correlation. On the other hand, a value of 0 indicates no relationship between the variables. This allows us to quickly understand the interconnections between the variables in our analysis. By examining the matrix, we easily identify which features have a high correlation with quality and are likely to be significant contributors to the machine learning models.

\begin{figure}
\begin{center}
\includegraphics[width=10cm, height=7cm]{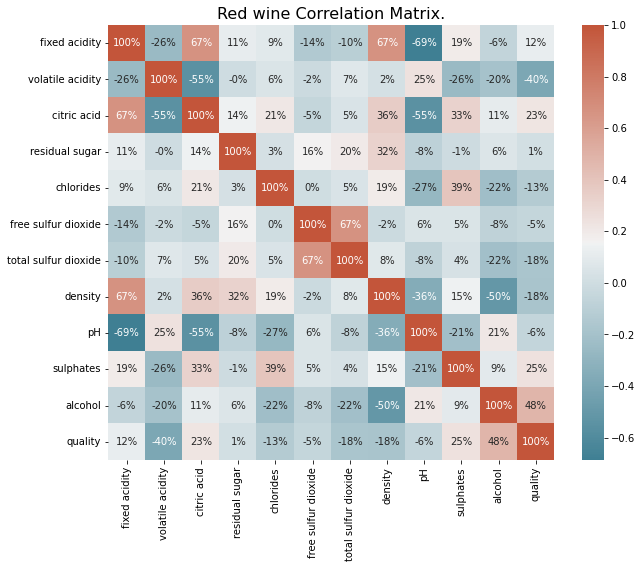}
\end{center}
\caption{Red wine correlation matrix}
\label{figs1}
\end{figure}

In Figure \ref{figs1}, we can see a correlation matrix showing a visual representation of the relationship between several variables, including "quality vs. alcohol," "volatile acidity vs. alcohol", "density vs. alcohol", and "sulphates vs. alcohol". Although the primary objective of this study is to identify features that are most indicative of good wine quality, it is evident from Figure \ref{figs1} that certain features such as alcohol, volatile acidity, and chlorides, exhibit the highest correlations with quality. This suggests that these variables have the most significant impact on predicting the quality of the wine.

The feature selection process aims to reduce the number of input variables in a machine learning model by identifying and retaining only the relevant data. This can be achieved by choosing the features that are likely to be useful in finding a solution to the problem, thereby reducing noise in the data and enhancing the performance of the model \cite{kira1992practical}. One of the objectives of this study is to look into the relationship between various features through the use of Pearson's correlation coefficient to quantify the associations between the different features.

In Table~\ref{tafile2} features are ranked according to their correlation values. 
According to \cite{liu2020daily} Pearson correlation coefficient $\rho$ given a pair of random variables $(X,Y)$ where $X$ and $Y$ are features, the formula for $\rho$ is 
\begin{equation}
\rho _{x,y} = \frac{cov(X,Y)}{\sigma _{X} \sigma _{Y}},
\end{equation} 
where $cov$ is the covariance, $\sigma _{X}$ is the standard deviation of feature $X$ and $\sigma _{Y}$ is the standard deviation of feature $Y$.

Table \ref{tafile2} presents the selected features, out of which 10 were chosen for further analysis. However, following the principle of selecting essential features for improved model performance as suggested by \cite{gupta2018selection}, we excluded 'residual sugar' based on our machine learning model's consistently better performance without it. This decision was supported by data indicating that 'residual sugar' had a relatively minor impact on wine quality compared to other variables. Figure \ref{ResidualVSquality} (shown below) visually illustrates the relationship between quality and residual sugar. It is observed that quality tends to increase when residual sugar is minimal and remains relatively unchanged beyond a certain point. This finding suggests that "residual sugar" is not as crucial as other variables such as alcohol in determining the quality of the wine. Figure \ref{AlcoholVSquality} depicts quality against alcohol, we can clearly see that alcohol is greatly contributing to the quality of wine, as the quality of wine increases we can see that the alcohol also increases. The results of the analysis revealed that the models performed better with the selected features as compared to when we used all the features.
\begin{table}
   \centering
   \begin{tabular}{|c|c|c|c|c|c|}
   \hline
   		\textbf{Rank} & \textbf{Name} & \textbf{Correlation} \\
   		\hline\hline
   		1 & alcohol & 48\%  \\
   		2& volatile acidity& -40\%\\
   		3 &sulphates & 25\%\\
   		4& citric acid& 23\%\\
   		5& total sulfur dioxide& -18\%\\
   		6& density& -18\%\\
   		7& chlorides& -13\%\\
   		8& fixed acidity& 12\%\\
   		9& pH& -6\%\\
   		10& free sulfur dioxide& -5\%\\
   		11& residual sugar& 1\%\\
   		\hline
   \end{tabular}
   \caption{Correlation with Quality}
   \label{tafile2}
\end{table}

Data standardization is a process that involves transforming data into a standardized form that will ensure that its distribution has a standard deviation of 1 and a mean of 0. The process of data standardization is essential as it helps in equalizing the range of information \cite{dahal2021prediction}, allowing for a more fair comparison between different features. For instance, as shown in Table \ref{tafile1} the overall Sulfur Dioxide readings are notably greater than chlorides. When we train machine learning models, having one variable with an exceptionally high value can mask all others, causing bias. Hence we need to standardize our data.
\begin{figure}
\begin{center}
\includegraphics[width=10cm, height=6cm]{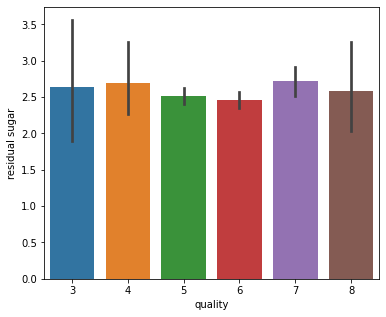}
\end{center}
\caption{Residual sugar versus quality}
\label{ResidualVSquality}
\end{figure}
\begin{figure}
\begin{center}
\includegraphics[width=10cm, height=6cm]{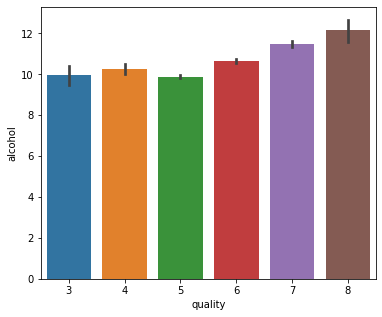}
\end{center}
\caption{Alcohol versus quality}
\label{AlcoholVSquality}
\end{figure}
\section{Classification Methods}
\subsection{Support Vector Machine}
SVM is one of the most well-known supervised learning algorithms that maximizes the margin. The goal of a support vector machine is to find a hyperplane that can efficiently separate various classes of data points within a high-dimensional space. This will enable us to swiftly classify new data points \cite{zhang2001maximal}. A hyperplane is the optimal decision boundary. The SVM algorithm takes into account the various extreme points that help in creating a hyperplane. The SVM algorithm is used for both linear (separable case) and non-linear (non-separable case) data. Let $D=\left\{(x_{i},y_{i})\right\}_{i=1}^{N}$ where $(x_{i},y_{i})$ represents an individual data point and its corresponding label, and $D \in \mathbb{R}^{m\times n}$ be a training data with $m$ rows and $n$ columns. Here $x_{i} \in \mathbb{R}^{n}$ and $y_{i} \in \left\{0,1,2\right\}$ are indicating a multi-class classification with 0 as bad quality wine, 1 as normal wine, and 2 as good quality wine. We construct a function to classify the quality of the wine based on its features $x_i$.  \\
\begin{gather*}
f : \mathbb{R}^{n} \to \mathbb{R}\\
x_{i}\mapsto f(x_{i})=
\begin{cases}
  0, \text{ if wine is bad quality, or}\\    
  1, \text{ if wine is normal quality, or}\\
  2, \text{ if wine is good quality}\\
\end{cases}
\end{gather*}
\subsubsection{Linear SVM (separable case)}
\hfill\vspace{0.3cm}\\
According to \cite{zhang2001maximal}, we first assume that the training data are linearly separable and that there is a hyperplane that separates the data without error. In this case, we look for the maximum margin hyperplane:
\begin{equation}
f(x)=\bigl \langle w,x\bigr \rangle + b = w^{T}x+b.
   \label{Max_M_Hyperplane}
\end{equation}\\
Where $\langle \cdot,\cdot\rangle$ and $w^{T}$ are the inner product and the transpose of the vector $w$ respectively. If $x_s$ is a support vector and $H=\bigl\{x|w^{T}x+b=0\bigr\},$ then the margin is given by:
\begin{equation}
\begin{aligned}
\text{Marge} &= 2d\bigl (x_s,H \bigr ) \\
&= \frac{2|w^{T}x+b|}{||w||},
\end{aligned}
\end{equation}

where $w$ is a normal vector called weight, $x$ is the input vector and $b$ is a bias.
The parameter $w$ and $b$ are not unique, and $kw$ and $kb$ give the same area of separation:
\begin{equation}
\begin{aligned}
kw^{T}x+kb &= k\bigl(w^{T}x+b\bigr) \\
&= 0.
\end{aligned}
\label{Marge_C}
\end{equation}

We then impose the normalization condition $\bigl |w^{T}x_s+b \bigr |=1$ for the $x_s$ support vectors, which leads to:
\begin{equation}
Marge=\frac{2}{||w||}.
   \label{Marge2}
\end{equation}\\
In order to minimize the margin, we thus need to minimize $||w||$. Recall the normalization conditions: $wx_i+b=1$ if $x_i$ is a support vector of class $+1$ and $wx_i+b=-1$ if $x_i$ is a support vector of class $-1$.:

\begin{gather*}  
\begin{cases}
  \text{if } y_i=1 \text{ then } wx_i+b\geq 1 \text{ and thus } y_i(wx_i+b) \geq 1\\    
  \text{if } y_i=-1 \text{ then } wx_i+b\le -1 \text{ and thus } y_i(wx_i+b) \geq 1\\
\end{cases}
\end{gather*}
We now must solve a quadratic programming problem of optimization (called primal problem):
\begin{gather*}  
\begin{cases}
  \text{min}_{w,b}\frac{1}{2}||w||^{2}\\    
  \text{if } y_i=-1 \text{ then } wx_i+b\le -1 \text{ and thus } y_i(wx_i+b) \geq 1\\
\end{cases}
\end{gather*}
The two parallel normal constraints of this optimization problem are separated by a Lagrange function. To solve this problem, we can combine the two constraints into a new Lagrangian function. We can also introduce new "slack variables" that denote $\alpha$ and require the derivative of the function to be zero. According to \cite{zhang2001maximal} the Lagrangian is given by:
\begin{equation}
L(w,b,\alpha)=\frac{1}{2}||w||^{2}+\sum_{i=1}^{n}\alpha _i \bigl[y_i\bigl(w^{T}x_i+b-1\bigr)\bigr],
\label{Lagrange}
\end{equation}
where $\alpha _{i}$ represents the Lagrange multiplier introduced to solve the constrained optimization problem.\\
\subsubsection{Linear SVM (Non-separable case)}
\hfill\vspace{0.3cm}\\
Hyperplane cannot completely segregate binary classes of data in the majority of real-world data, hence we accept some observations in the training data on the incorrect side of the margin or hyperplane. Here, is the primal optimization problem of Soft Margin:

\begin{gather*} 
\begin{cases}
  \min_{w,b}\bigl (\frac{1}{2}||w||^{2}-C\sum _{i=i}^{n}\xi_{i}\bigr )\\    
  y_i(wx_i+b) \geq 1-\xi_{i} \text{ and }\xi_{i}\geq 0, i=1,\cdots, n\\
\end{cases}
\end{gather*}
where $\xi_{i}$ is the slack variable that allows misclassification; the penalty term $\sum_{i}^{n}\xi_{i}$ is a measure of the total number of misclassification in the model and $C$ is a penalty variable for misclassified points \cite{LibreTexts}. Using the same terminology for separable SVM, we get the dual problem:
\begin{gather*} 
\begin{cases}
  \max_{\alpha}\bigl (\sum_{i=1}^{n} \alpha_{i} -\frac{1}{2}\sum_{i,j=1}^{n} \alpha_{i}\alpha_{j}y_{i}y_{j}(x_{i}x_{j})\bigr )\\    
  \sum_{i=1}^{n} \alpha_{i}y_{i}=0\\
  C\geq \alpha_{i} \geq 0 \text{, } i=1,\cdots, n.
\end{cases}
\end{gather*}

The classification of a new observation $x$ is determined by the decision function:

\begin{equation}
f(x)= \sum_{i=1}^{n} \alpha_{i}y_{i}(x_{i}x)+b.
\end{equation}
\subsection{Decision Tree}
A decision tree is a type of machine learning that takes into account the various inputs and outputs in a given training program. It then continuously splits the data according to a set of parameters. The two entities that comprise a decision tree are the leaves and the decision nodes \cite{Nagesh2022}.

Getting the correct attribute for a particular tree's root node is a huge challenge. This is why it is important to consider the various methods that are available to select attributes. There are two main methods that are commonly used to select attributes which are entropy and information gain. Let S be a sample and $S_1,\cdots, S_k$ the partition of $S$ according to the classes of the target attribute. The Gini is denoted as $Gini(S)$ and the entropy is denoted as $Ent(S)$ are defined by \cite{du2002building}.\\

\begin{equation}
Gini(S)=  \sum_{i=1}^{k}\frac{|S_i|}{|S|}\times\left(1-\frac{|S_i|}{S}\right)=\sum_{i\neq j}^{}\frac{|S_i||S_j|}{|S|},
   \label{Gini}
\end{equation}
\begin{equation}
\begin{aligned}
&\text{and the entropy as:}\\
&Ent(S)= -\sum_{i=1}^{k}\frac{|S_i|}{|S|}\times\log\left(\frac{|S_i|}{|S|}\right),
\end{aligned}
\label{Entropy}
\end{equation}
where $|S_{i}|$ is the cardinality in the set $S_{i}$ and $|S|$ is the cardinality in the sample $S$. The variables $i$,$ j$, and $k$ represent indices where $i$ refers to attribute classes, $j$ indicates different classes for the Gini formula, and $k$ represents the total class.
\subsection{Random Forest}
\begin{figure}
\begin{center}
\includegraphics[width=10cm, height=6cm]{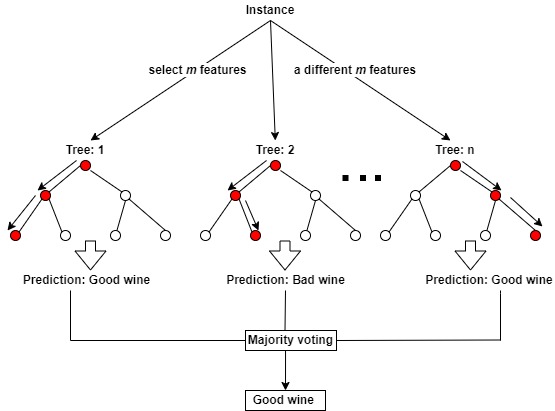}
\end{center}
\caption{ Random Forest(adapted from \cite{MaxKjell})}
\label{RandF}
\end{figure}

RF is a widely used algorithm that is a part of the supervised learning framework. It can be utilized for regression and classification problems. It's based on the idea of ensemble learning, in which multiple classifiers are combined to solve a given problem and to enhance the model's performance.\cite{Nagesh2022}. Figure \ref{RandF} demonstrates how random forest predicts the quality of the wine.

The RF classifier combines the power of numerous decision trees. It creates several decision trees using bootstrapped datasets and randomly chooses a subset of the variables for each stage. Figure~\ref{RandF} shows how RF works. It aggregates the predicted outcomes from all the decision trees, and it chooses the mode that is most likely to perform well. This approach ensures that the model is more accurate and reliable, minimizing the risk that a single tree can make an error. By adopting a "majority wins" approach, RF ensures that the ultimate prediction is derived from a collective agreement among the decision trees, instead of relying solely on the outcome of an individual tree.
\subsection{Gradient Boosting}
A Gradient Boosting Machine is a type of tool that creates a strong learner by taking weak individuals and merging them into a single model. It can be used for classification and regression tasks. Although it is mainly utilized for tree-based models, it can also be applied to other weak individuals \cite{natekin2013gradient}.

The fundamental concept behind GB involves incorporating new models into the ensemble, with each new model focusing on the examples that were incorrectly classified by the previous models. In order to focus on these difficult examples, GB fits each new model to the negative gradient of the loss function with respect to the current ensemble model \cite{Saini2021}. The GB method can be used in various applications such as regression, ranking problems, and classification.
\subsection{K-Nearest Neighbours}
\label{knn}
The KNN classifier is a machine learning algorithm used for classification and regression tasks that work on the premise that similar objects are usually located near each other \cite{LibreTexts}. In order for KNN to find the neighbors of a query point we need to calculate the distance between the query point and the other data points. These distance measures help in the formation of decision boundaries, which divide query points into distinct areas. One of the main drawbacks of the KNN algorithm is that it may be biased towards the majority class in datasets that are imbalanced, meaning that there are significantly more instances in one class than in another \cite{liu2011class}. This is because KNN classifies query points by finding the $k$ nearest neighbours in the training set and if the majority class dominates the neighbourhood of the test instance it is likely to be classified as the majority class.

Let's say we have a dataset with $X$ representing a matrix that contains the features observed and $Y$ representing the class label. Lets assume we have a point $x$ which has coordinates $(x_1, x_2,\cdots, x_p)$ and point $y$  with coordinates $(y_1, y_2,\cdots, y_p)$  \cite{LibreTexts}. The KNN algorithm is in this study because it categorizes new cases based on the Euclidean distance between the training data and the test observation. In k-NN, the optimal choice is determined by identifying the set of training data points that are closest to the given test observation in terms of Euclidean distance \cite{tamamadin2022regional}. 
\begin{equation}
d(x_{i},x_{t})=\sqrt{\sum_{j=1}^{d} \bigl (x_{ij}-x_{tj}\bigr)^2},\\
= \lVert x_{i}-x_{t} \rVert
\label{Distance}
\end{equation}
where $x_{i}$ represent the training data and $x_{t}$ represent the test observation.\\
Majority voting is the process of selecting the class that has the highest number of votes among the k-nearest neighbours in the K-nearest neighbours (KNN) algorithm. Majority voting is defined as follows according to \cite{liu2011class}:
\begin{equation}
\hat{f}(x_{t}) = \underset{c\in \{c_{1},c_{2},c_{3}\}}{\mathrm{argmax}}\text{ } \sum_{(x_{i},y_{i})\in N_{k}(x_{t})}I(y_{i}=c),
\label{argmax}
\end{equation}
where $x_{t}$ represent the test observation, $\hat{f}(x_{t})$ represent a forecasted class label, $N_{k}(x_{t})$ represent a set of training instances and I$(\cdot)$ represent an indicator function that takes a value as input and returns either 0 or 1 based on whether the input satisfies a certain condition\cite{liu2011class}.
\begin{figure}
\begin{center}
\includegraphics[width=10cm, height=6cm]{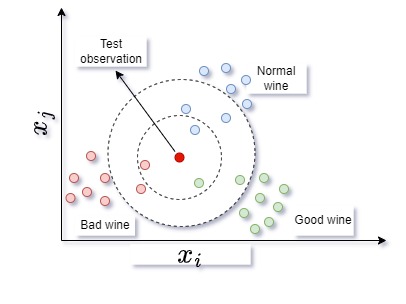}
\end{center}
\caption{KNN with different k-values (adapted from\cite{LibreTexts})}
\label{KNN_fig}
\end{figure}

Figure~\ref{KNN_fig} shows the KNN classifier with K= 3 and K= 7. We need to predict the class for the new observation (red circle) if it belongs to a class of Bad wine, Normal wine, or Class of Good wine respectively. If we choose k=3 (for a small dotted circle) then we have one observation in Class Bad wine, one observation in Class Normal wine, and one observation in Class Good wine. From this we have Pr(Bad wine)=$\frac{1}{3}$, Pr(Normal wine)=$\frac{1}{3}$ and Pr(Good wine)=$\frac{1}{3}$ respectively. We can clearly see that we have a tie among our classes where each class has one observation. Since the number of neighbours in class Bad Wine, class Normal, and class Good Wine are the same, we cannot determine the class of the new data point based on the number of neighbours alone. According to \cite{Ashok2020}, we can use different tie-breaking techniques to determine the class in case of a tie. One common method is to choose the class that has the shortest average distance to the new data point. If we choose  k=7 (for a big dotted circle) then we have two observations in Class Bad Wine, three observations in Class Normal Wine, and two observations in Class Good Wine. From this we have Pr(Bad wine)=$\frac{2}{7}$, Pr(Normal wine)=$\frac{3}{7}$ and Pr(Good wine)=$\frac{2}{7}$, so we can clearly see that the small red circle (test observation) belongs to class Normal wine based since class Normal wine has the highest probability as compared to other classes (majority voting). The value of a classier determines the performance of that class. However, selecting the correct different $k$ values can be very challenging. This is because the value of $k$ can have a huge impact on the accuracy of the predictive model \cite{ling2021improved}.\\

\section{ Experimental Settings}
\subsection{Unbalanced Data}
\begin{figure}
\begin{center}
\includegraphics[width=10cm, height=6cm]{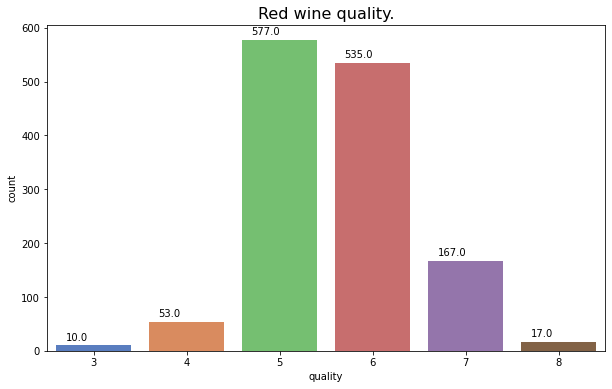}
\end{center}
\caption{Distribution of red wine quality}
\label{fig3}
\end{figure}

Figure \ref{fig3} demonstrates the red wine quality classes, the dataset's distribution can be seen with the most significant value being 5 with the class values ranging from 3 to 8.

The dataset depicts an unbalanced distribution of red wine with other classes not being fairly represented, the instances range from 10 in the minority class to 577. As suggested by \cite{chawla2009data}, sampling techniques such as undersampling, oversampling, and SMOTE are used to handle unbalanced datasets. These are further discussed in section \ref{B}.\\

\subsection{Sampling Techniques}{\label{B}}
\subsubsection{Undersampling and Oversampling}
\hfill\vspace{0.3cm}\\
The oversampling method is an intuitive technique that increases the size of a minority class by creating duplicates of samples taken from the under-represented group. Undersampling on the other hand ensures that all of the data from the minority segment are kept and reduces the size of the majority segment to be the same as the minority segment. Undersampling is usually considered to be a disadvantage as it eliminates potentially useful data. Oversampling on the other hand is more likely to cause overfitting since it duplicates existing examples \cite{shelke2017review}.
\subsubsection{Synthetic Minority Oversampling Technique}
\hfill\vspace{0.3cm}\\
According to \cite{chawla2009data}, using the SMOTE filter proves to be a valuable approach in addressing imbalanced wine datasets. SMOTE employs a k-nearest neighbour method to create synthetic data points. SMOTE starts by selecting K nearest neighbours from the minority samples based on the desired level of oversampling and then randomly selecting a neighbour from K nearest neighbors\cite{chawla2009data}. The selection process is not deterministic as the K nearest neighbours are chosen randomly. The selection of the K neighbors is done randomly and the random data is combined to generate synthetic data. SMOTE utilizes synthetic data points to add diversity to the minority class, mitigating the issue of overfitting that arises from random sampling techniques. According to \cite{chawla2009data} SMOTE also creates a more balanced dataset which can help improve the performance of machine learning models when dealing with imbalanced data.

\subsection{Hyper parameter tuning}
In machine learning, the task of selecting a set of optimal hyperparameters for a learning algorithm is known as hyperparameter tuning.
The simplest approach to tuning hyperparameters is undoubtedly grid search. Using this method, we simply construct a model for every possible combination of the supplied hyperparameter values and evaluate each model, and choose the model that yields the best results \cite{zahedi2021search}. According to \cite{wu2019hyperparameter}, hyperparameter optimization is expressed as:
\begin{equation}
x^{*}=\underset{x\in X}{\arg\min} f(x),
\end{equation}
where $f(x)$ represents a score that we aim to minimize, such as the error rate evaluated on the validation set. $x^{*}$ refers to the set of hyperparameters that produces the lowest score value while $x$ can take any value within the $X$ domain.
With this, we want to determine the model hyperparameters that provide the highest score on the validation set metric.
\subsection{Model Evaluation}
To understand how well and efficiently the model performs, we measure and evaluate its performance. There are four techniques used to determine the accuracy of predictions:
\begin{itemize}

    \item True Positive (TP): This indicates the percentage of samples that the model correctly identifies as positive.
    \item False Positive (FP): It represents the percentage of samples that the model mistakenly predicts as positive when they are actually negative.
    \item False Negative (FN): These are the samples that the model wrongly classifies as negative while they are positive in reality.
    \item True Negative (TN): These are the samples that the model accurately identifies as negative.
\end{itemize}

We use the following techniques to assess the model.
\begin{enumerate}
    \item Accuracy: It can be characterized as either the proportion of all positive classes that the model correctly predicted to be true or the number of accurate outputs that the model provides. Its formula is:
\begin{equation}
Accuracy= \frac{TP + TN}{TP + TN + FP + FN}.
   \label{Accuracy}
\end{equation}\\
    \item Precision: Precision refers to the ratio of predicted observations to the total number of expected positive observations. Its formula is:
\begin{equation}
Precision= \frac{TP}{TP + FP}.
   \label{Precision}
\end{equation}\\
    \item Recall: Recall is known as the proportion of accurately predicted positive observations to all of the actual class observations. Its formula is:
\begin{equation}
Recall= \frac{TP}{TP + FN}.
   \label{Recall}
\end{equation}\\
    \item $F_1$ Score: $F_1$ score is calculated as the balanced average of recall and accuracy. The test accuracy of the model is evaluated using the $F_1$ score. Its formula is \cite{kumar2020red}:
\begin{equation}
F_1 Score= 2\times\frac{Recall\times Precision}{Recall + Precision}.
   \label{F1 Score}
\end{equation}\\
\end{enumerate}
According to \cite{guo2008class}, accuracy is the primary metric used to evaluate models, but when dealing with skewed class distributions and imbalanced datasets, it becomes challenging to make accurate judgments. For instance, the recall rate for minority groups has typically dropped to zero. This indicates that the model is not able to properly classify them. The reduction in recall and precision scores for minority groups is due to how the model focuses more on the majority segment instead of the minority segments. This issue is caused by the preference of the accuracy model for the majority group. As a result, the classifier tends to perform poorly on the minority groups.

\section{Results and Discussion}
\subsection{Results}
For the purpose of this study, we are using five machine learning algorithms to predict the wine quality namely SVM, DT, KNN, GB, and RF. We implemented our models in an unbalanced dataset with default parameters and the results are shown in Table \ref{newT} below that our models performed poorly with support vector machine and random forest having the highest accuracy with 78\% each. Table \ref{newT} provides a comprehensive overview of our models' performance across metrics such, as accuracy, precision, recall, and F1 score.  

\begin{table}
\begin{center}
\includegraphics[width=10cm, height=4cm]{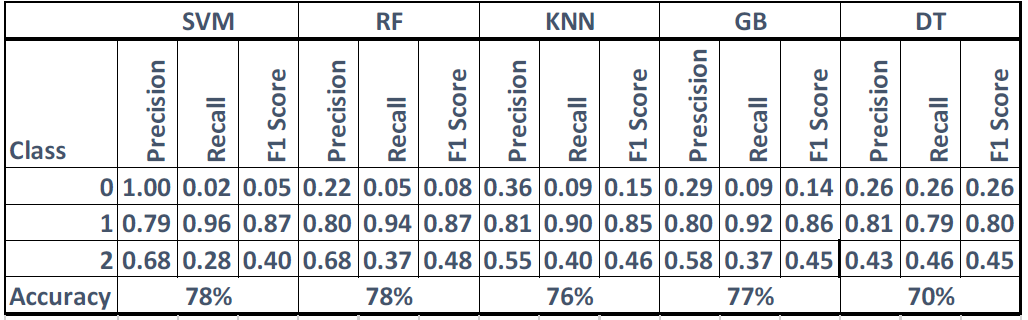}
\end{center}
\caption{Test results for the unbalanced dataset with default model parameters}
\label{newT}
\end{table}

We also implemented our models on a balanced dataset with tuned parameters. The results are shown in Table \ref{newt}, indicating that the models perform well compared to when the models were implemented in an unbalanced dataset with default parameters. As shown in Table \ref{newt} among the five machine learning algorithms used in this research to predict wine quality, SVM shows the best performance. As mentioned in section~\ref{knn} the KNN classifier in an unbalanced dataset tends to favour the majority class, this is evident in Table \ref{newT} as we can see that the precision, recall, and F1-score are high in the majority class (Class 1) as compared to other classes (Class 0 and Class 2). We can see that balancing the data and tuning your models increase the performance of your models as suggested by \cite{chawla2009data}. This is evident in Table \ref{newt} as we can see that the accuracy of our models increases as compared to when they were implemented our model in an unbalanced dataset with default parameters.
\begin{table}
\begin{center}
\includegraphics[width=10cm, height=4cm]{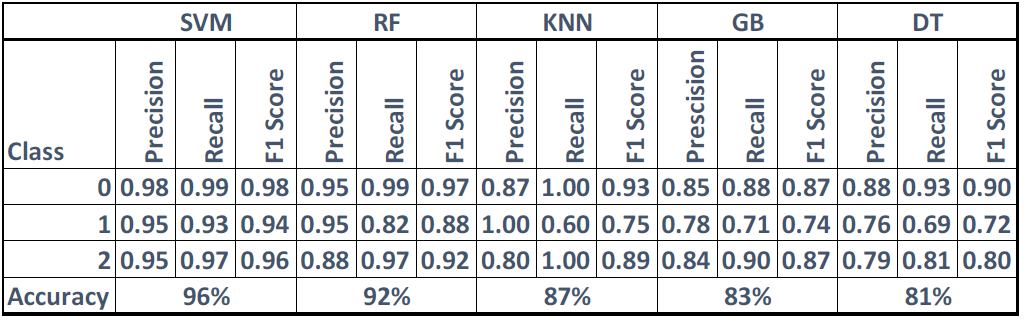}
\end{center}
\caption{Test results on the balanced dataset with tuned model parameters}
\label{newt}
\end{table}

\subsection{Feature importance}
\begin{figure}
\begin{center}
\includegraphics[width=10cm, height=6cm]{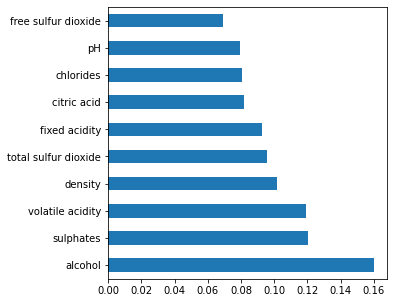}
\end{center}
\caption{Feature importance for our best performing model}
\label{newFeature}
\end{figure}

We also graphed the feature importance based on our best-performing machine learning model which is in this case the SVM. As we can see the feature importance graphed in Figure \ref{newFeature} alcohol is the most significant factor impacting wine quality, and this was also suggested by \cite{mor2022wine} that alcohol plays a crucial role in determining wine quality. Looking at the feature importance graph it suggests that tuning features such as "alcohol", "sulphates", and "volatile acidity" may increase or decrease the wine scores. This information suggests that winemakers may benefit from tuning their models and playing around with the physio-chemical properties of wine.

\subsection{Discussion}
The objective of this research is to try and predict the quality of wine by analyzing the physico-chemical properties of the wine. It also looks into which features of the wine are most indicative of its quality. To achieve this goal we applied several machine learning algorithms as mentioned above, including Random Forest (RF), Support Vector Machine (SVM), Gradient Boosting (GB), K-Nearest Neighbors (KNN), and Decision Tree (DT). We chose to use these machine learning algorithms because they are widely used algorithms for classification problems and are effective for wine quality prediction. We also dug deeper into the data and we found an interesting relationship between our feature variables and the target variable (Quality). We used the correlation coefficient matrix as shown in Figure 1 and features are ranked according to their correlation values. The results shown in Figure 1 suggest that features like "alcohol", "volatile acidity", and "sulphates" have a high correlation with quality while features like "free sulfur dioxide" and "residual sugar" do not. In Table  \ref{tafile2}, features are ranked according to their correlation values, and the first 10 features are selected during the models' ultimate implementation.\\    

We assessed the effectiveness of the algorithms by analyzing metrics, including precision, recall, accuracy, and $F_1$ score as presented in both Table \ref{newT} and Table \ref{newt}. We then evaluated the performance of our model by applying it to both the imbalanced dataset with default parameters and to a balanced dataset with fine-tuned parameters. The results of the analysis are presented in Table \ref{newT} and Table \ref{newt} respectively. From the performance results it is evident that the best outcome is achieved with a balanced dataset with fine-tuned parameters. As mentioned above it is evident that balancing the data and tuning your model parameters enhances the models' performance.\\ 

\section{Conclusion}
This study showed the importance of feature selection
in understanding the impact of analytical tests on wine
quality. The results of the feature selection process showed
that some input variables such as Alcohol had a more significant influence on predicting wine quality than others such as Residual sugar. Applying machine learning algorithms in conjunction with the results of the feature selection process presented a valuable opportunity to improve the wine production process. \\

We employed five machine learning models, namely Decision Tree (DT), Random Forest (RF), Support Vector Machine (SVM), K-Nearest Neighbors (KNN), and Gradient Boosting (GB). The Support Vector Machine (SVM) outperformed the other models with an accuracy of 96\%. Therefore, we conclude that not all features were equally important for predicting wine quality and that tuning your models and balancing the dataset improved the performance of the models. We also saw that in Figure \ref{newFeature} our feature importance graph suggested that tuning the models and playing around with physio-chemical properties such as "Alcohol" and "sulphate" may be beneficial in improving the prediction of wine quality.

Although this study presents promising results in predicting the wine quality using machine learning algorithms some limitations need to be addressed in future work, such as the small size of the dataset and we did not use all the algorithms. In future work using larger and more diverse datasets could enhance the machine learning algorithm's performance. This will help the algorithms generalize better and reduce the risk of overfitting, thus improving the wine production process. For our study, we only used Five machine learning algorithms and there are still many other algorithms that could be explored in future work. We can evaluate the performance of the algorithms using different metrics and we can explore the impact of different preprocessing techniques such as different feature scaling techniques on the performance of the algorithms.
\bibliographystyle{unsrt}
\bibliography{samplepaper}

\begin{thebibliography}{10}

\bibitem{adam2009health}
Bal{\'a}zs {\'A}d{\'a}m, {\'A}gnes Moln{\'a}r, Helga B{\'a}rdos, and R{\'o}za
  {\'A}d{\'a}ny.
\newblock Health impact assessment of quality wine production in hungary.
\newblock {\em Health promotion international}, 24(4):383--393, 2009.

\bibitem{saremi2008cardiovascular}
Adonis Saremi and Rohit Arora.
\newblock The cardiovascular implications of alcohol and red wine.
\newblock {\em American journal of therapeutics}, 15(3):265--277, 2008.

\bibitem{Meyer2019}
Meyer M.
\newblock The subtle science of wine tasting.
\newblock \url{https://winefolly.com/deep-dive/science-of-wine-tasting/}, 2019.

\bibitem{dahal2021prediction}
KR~Dahal, JN~Dahal, H~Banjade, and S~Gaire.
\newblock Prediction of wine quality using machine learning algorithms.
\newblock {\em Open Journal of Statistics}, 11(2):278--289, 2021.

\bibitem{Dua:2019}
Dheeru Dua and Casey Graff.
\newblock {UCI} machine learning repository, 2017.

\bibitem{gupta2018selection}
Yogesh Gupta.
\newblock Selection of important features and predicting wine quality using
  machine learning techniques.
\newblock {\em Procedia Computer Science}, 125:305--312, 2018.

\bibitem{pawar2019wine}
Devika Pawar, Aakanksha Mahajan, Sachin Bhoithe, M~Prasanna, and Kamalesh
  Kumar.
\newblock Wine quality prediction using machine learning algorithms.
\newblock {\em International Journal of Computer Applications Technology and
  Research}, 8(9):385--388, 2019.

\bibitem{cortez2009modeling}
Paulo Cortez, Ant{\'o}nio Cerdeira, Fernando Almeida, Telmo Matos, and Jos{\'e}
  Reis.
\newblock Modeling wine preferences by data mining from physicochemical
  properties.
\newblock {\em Decision support systems}, 47(4):547--553, 2009.

\bibitem{kira1992practical}
Kenji Kira and Larry~A Rendell.
\newblock A practical approach to feature selection.
\newblock In {\em Machine learning proceedings 1992}, pages 249--256. Elsevier,
  1992.

\bibitem{liu2020daily}
Yaqing Liu, Yong Mu, Keyu Chen, Yiming Li, and Jinghuan Guo.
\newblock Daily activity feature selection in smart homes based on pearson
  correlation coefficient.
\newblock {\em Neural Processing Letters}, 51(2):1771--1787, 2020.

\bibitem{zhang2001maximal}
Bin Zhang.
\newblock Is the maximal margin hyperplane special in a feature space.
\newblock {\em Hewlett-Packard Research Laboratories Palo Alto}, 2001.

\bibitem{LibreTexts}
LibreTexts.
\newblock K nearest neighbors.
\newblock
  \url{https://stats.libretexts.org/Bookshelves/Computing_and_Modeling/RTG%3A_Classification_Methods/3%3A_K-Nearest_Neighbors_(KNN)},
  2020.
\newblock Accessed in 25 August 2022.

\bibitem{Nagesh2022}
Nagesh~Singh Chauhan.
\newblock Random forest vs decision tree: Key differences.
\newblock
  \url{https://www.kdnuggets.com/2022/02/random-forest-decision-tree-key-differences.html},
  2022.
\newblock Accessed in 25 August 2022.

\bibitem{du2002building}
Wenliang Du and Zhijun Zhan.
\newblock Building decision tree classifier on private data.
\newblock 2002.

\bibitem{MaxKjell}
Kjell~Johnson Max~Kuhn.
\newblock Comparison analysis of machine learning algorithms: Random forest and
  catboost.
\newblock
  \url{https://rstudio-pubs-static.s3.amazonaws.com/740098_4d48bd29722f402abf662dd33fc67794.html},
  2020.
\newblock Accessed in 25 August 2022.

\bibitem{natekin2013gradient}
Alexey Natekin and Alois Knoll.
\newblock Gradient boosting machines, a tutorial.
\newblock {\em Frontiers in neurorobotics}, 7:21, 2013.

\bibitem{Saini2021}
Anshul Saini.
\newblock Gradient boosting algorithm: A complete guide for beginners.
\newblock
  \url{https://www.analyticsvidhya.com/blog/2021/09/gradient-boosting-algorithm-a-complete-guide-for-beginners/},
  2021.
\newblock Accessed in 03 June 2022.

\bibitem{liu2011class}
Wei Liu and Sanjay Chawla.
\newblock Class confidence weighted k nn algorithms for imbalanced data sets.
\newblock In {\em Advances in Knowledge Discovery and Data Mining: 15th
  Pacific-Asia Conference, PAKDD 2011, Shenzhen, China, May 24-27, 2011,
  Proceedings, Part II 15}, pages 345--356. Springer, 2011.

\bibitem{tamamadin2022regional}
Mamad Tamamadin, Changkye Lee, Seong-Hoon Kee, and Jurng-Jae Yee.
\newblock Regional typhoon track prediction using ensemble k-nearest neighbor
  machine learning in the gis environment.
\newblock {\em Remote Sensing}, 14(21):5292, 2022.

\bibitem{Ashok2020}
Ashok Reddy.
\newblock K nearest neighbors conceptual understanding and implementation in
  python.
\newblock
  \url{https://www.citrusconsulting.com/k-nearest-neighbors-conceptual-understanding-and-implementation-in-python/},
  2020.

\bibitem{ling2021improved}
YuLong Ling, Xiao Zhang, and Yong Zhang.
\newblock Improved knn algorithm based on probability and adaptive k value.
\newblock In {\em 2021 7th International Conference on Computing and Data
  Engineering}, pages 34--40, 2021.

\bibitem{chawla2009data}
Nitesh~V Chawla.
\newblock Data mining for imbalanced datasets: An overview.
\newblock {\em Data mining and knowledge discovery handbook}, pages 875--886,
  2009.

\bibitem{shelke2017review}
Mayuri~S Shelke, Prashant~R Deshmukh, and Vijaya~K Shandilya.
\newblock A review on imbalanced data handling using undersampling and
  oversampling technique.
\newblock {\em Int. J. Recent Trends Eng. Res}, 3(4):444--449, 2017.

\bibitem{zahedi2021search}
Leila Zahedi, Farid~Ghareh Mohammadi, Shabnam Rezapour, Matthew~W Ohland, and
  M~Hadi Amini.
\newblock Search algorithms for automated hyper-parameter tuning.
\newblock {\em arXiv preprint arXiv:2104.14677}, 2021.

\bibitem{wu2019hyperparameter}
Jia Wu, Xiu-Yun Chen, Hao Zhang, Li-Dong Xiong, Hang Lei, and Si-Hao Deng.
\newblock Hyperparameter optimization for machine learning models based on
  bayesian optimization.
\newblock {\em Journal of Electronic Science and Technology}, 17(1):26--40,
  2019.

\bibitem{kumar2020red}
Kanika Kumar and Nelshan Mandan.
\newblock Red wine quality prediction using machine learning techniques.
\newblock In {\em 2020 International Conference on Computer Communication and
  Informatics (ICCCI)}, pages 1--6. IEEE, 2020.

\bibitem{guo2008class}
Xinjian Guo, Yilong Yin, Cailing Dong, Gongping Yang, and Guangtong Zhou.
\newblock On the class imbalance problem.
\newblock In {\em 2008 Fourth international conference on natural computation},
  volume~4, pages 192--201. IEEE, 2008.

\bibitem{mor2022wine}
Nuriel~Shalom Mor.
\newblock Wine quality and type prediction from physicochemical properties
  using neural networks for machine learning: A free software for winemakers
  and customers.
\newblock 2022.

\end{thebibliography}
\end{document}